\documentstyle[multicol,aps,prl,epsf]{revtex}
\begin{document}

\draft

\title{Electronic properties of alkali-metal loaded zeolites \\
--- a ``supercrystal" Mott insulator
} 

\author
{Ryotaro~Arita$^1$, Takashi~Miyake$^2$, Takao~Kotani$^3$, 
Mark~van~Schilfgaarde$^4$, Takashi~Oka$^1$, Kazuhiko~Kuroki$^5$, 
Yasuo~Nozue$^3$ and Hideo~Aoki$^1$}

\address{$^{1}$Department of Physics, University of Tokyo,
Hongo, Tokyo 113-0033, Japan}
\address{$^{2}$Department of Physics, Tokyo Institute of Technology,
Oh-okayama, Tokyo 152-8551, Japan}
\address{$^{3}$Department of Physics, Osaka University,
Toyonaka, Osaka 560-0043, Japan}
\address{$^{4}$Department of Chemical and Materials Engineering,
Arizona State University,Tempe, Arizona 85287-6006, USA}
\address{$^{5}$Department of Applied Physics and Chemistry,
University of Electro-Communications, Chofu, Tokyo 182-8585, Japan}

\date{\today}
\maketitle

\begin{abstract}
First-principles band calculations are performed for the first
time for an open-structured zeolite (LTA) with guest atoms 
(potassium) introduced in
their cages.  A surprisingly simple band structure emerges, which
indicates that this system may be regarded as a
``supercrystal", where each cluster of guest atoms with diameter
$\sim$10\AA\ acts as a ``superatom" with well-defined $s$- and
$p$-like orbitals, which in turn form the bands around the Fermi energy.  
The calculated Coulomb and exchange energies for 
these states turn out to be in the strongly-correlated regime.  
With the dynamical mean-field theory we show the system 
should be on the Mott-insulator side, and, on a magnetic 
phase diagram for degenerate-orbital systems, around the 
ferromagnetic regime, in accord with experimental results.
We envisage this class of systems can provide a
new avenue for materials design.
\end{abstract}

\medskip


\begin{multicols}{2}
\narrowtext
Materials design is usually done 
by varying chemical elements and/or atomic crystal
structures.  
Here we consider doped zeolites, which 
are unique in that (i) they are
host-guest systems, where the host itself appears in a rich
variety (crystal structure of cages, size of the
inter-cage window, etc), on top of rich possibilities for the
species and the number of guest atoms\cite{Bogomolv78}. 
(ii) the size of the cage is $\sim$ nanometer.  
We can note the distinction of the zeolites from the solid fullerene: The
latter are also nanostructured supercrystals, but the relevant
orbitals, even when doped with alkali metals, 
are LUMO/HOMO of the cage (buckeyball) 
rather than those of dopants\cite{saitooshiyama}.  
While (i) implies higher degrees of freedom for the design, (ii)
implies that the Coulomb interaction energy for the cluster doped
in the cage, which only decreases inversely with the size, can remain
large ($\sim$ eV for nanometer dimensions).  Here we 
envisage that these can provide a fascinating 
playing ground for {\em systematically control} the electron
correlation as a new avenue for materials
design.  

Indeed, experimental results by Nozue {\em et al} showed some
zeolites (Fig.1) loaded with clusters ($\simeq$ five atoms
per cage) of potassium is ferromagnetic for
$T<8$K\cite{Nozue92,Nozue93}, which is unique in that magnetism
occurs even though all the ingredients are non-magnetic
elements\cite{commentmag}.  At the same time, an infrared
analysis for the K-loaded zeolite shows clear absorption edges,
indicative of insulating behavior\cite{Nakano99}.

Here we address both phenomena, and show they can be 
explained in terms of strong electron correlations 
in a surprisingly simple electronic structure.  
We start by showing that the relevant states
of K-doped zeolite can be viewed from a ``supercrystal" point of
view.  Namely, the single-particle electronic structure 
as computed within the local-density approximation (LDA) 
for a typical zeolite having a cubic array of cages 
is very accurately described as a tight-binding band of
``superatoms" (i.e., the states in the clusters of dopants) 
with simple $s$ and $p$ orbitals.  This explains why
the optical spectrum can be interpreted as transitions between
$s$ and $p$ (or $p$ and $d$) orbitals in a simplified 
well\cite{Kodaira93}.  
For the LDA calculation we adopt the all-electron
full-potential linear muffin-tin orbitals (FP-LMTO)\cite{Mark}.  In fact
ours is the first reliable band structure calculation for zeolites\cite{Sun98}. 

Next we estimate the Coulomb and exchange
interactions for these orbitals by plugging in the 
LDA wave functions, and show that the interactions are large
compared with the single-particle bandwidth, indicating strong
electron correlations.  Last, we study the effect of correlations
on the spectral function in the spirit of the LDA+dynamical
mean-field theory \cite{lda+dmft}, and show the system resides well 
on the Mott-insulator side.  This resolves a puzzle that the band calculation
predicts that the system is a metal, while experiments indicate
an insulator; moreover we show the calculated exchange energy (Hund's
coupling in superatoms) indicates ferromagnetism 
in the multi-orbital magnetic phase diagram, in accord with experiments.

We start with the undoped zeolite with a typical structure 
having a simple-cubic array of cages (Fig.1), where each cage
(called $\alpha$) is an Archimedes polyhedron. The region
surrounded by eight $\alpha$ cages forms another cage called
$\beta$.  Aluminosilicate with this structure (zeolite A) has Si
and Al atoms situated at the vertices of the cages and connected
by O atoms with the inner diameter of $\alpha$ cage $\sim 10$
{\AA}.
The material used in most experiments\cite{Nozue92,Nozue93} is
K(potassium)-form 
zeolite A (abbreviated as LTA hereafter), whose chemical formula, 
K$_{12}$Al$_{12}$Si$_{12}$O$_{48}$ (with 84 atoms per unit cell), 
already contains some K atoms.  Eleven out of the twelve K atoms 
are located on the centers of faces of the $\alpha$ cage 
(orange atoms in Fig.1), 
while the remaining K (red), assumed to be at the cage center, 
changes its position as K's are added to change the 
doping as we shall describe.
For the atomic positions of Al, Si, O within the unit cell, we have 
adopted the accurate crystal structure obtained from a recent 
neutron powder diffraction study by Ikeda {\it et al}\cite{Ikeda00,Lowenstein}.

The result for the band structure 
for the undoped LTA is displayed in Fig.2, which 
shows that the undoped zeolite is an 
insulator with a gap $\simeq 4$ eV, where the 
wave functions indicate that, 
while the valence-top state sits on the framework, the 
conduction-band states (labeled as B-F) primarily 
reside within the $\alpha$ cage (for B,D,E,F) or within $\beta$ (C).  

The dispersion of the lowest conduction band (B) can be fitted, 
as shown in the right panel of Fig.2, by the dispersion
of a tight-binding band on the simple cubic lattice, 
$\varepsilon({\bf k})=t_x \cos(k_x)+t_y \cos(k_y)+t_z \cos(k_z)$,
with ($t_x,t_y,t_z) \simeq (-20, -10, -5)$ (meV).
The fitting is excellent, and, with all of $t_x$, $t_y$, $t_z < 0$ 
and the wave function being nodeless (Fig.2(B), with its tetrahedral shape 
due to the configuration of surrounding K atoms), 
we may interpret the band as an ``$s$ band" of the supercrystal. 
Similarly, three (D, E, F) out of the 
four next-lowest conduction bands, with large amplitudes within the
$\alpha$ cage, may be interpreted as $p$ bands.

Now we come to the doped case. When one K atom is doped 
per unit cell (denote as K$_1$LTA), the Fermi level shifts to the middle 
of the conduction ($s$) band (not shown), while the $p$ band remains empty.  
Tight-binding fits to these bands are again excellent, with 
$(t_x, t_y, t_z) \simeq (-10, -10, -5)$ in meV for the $s$ band and 
$\simeq (25, -10, -1)$ for the $p$. The good fit to the simple 
tight-binding model is highly nontrivial, since the framework, 
(K$^+)_{12}$(Al$^{3+})_{12}$(Si$^{4+})_{12}$O$^{2-})_{48}$ 
with nominal valence indicated, is an ionic compound.  
Chemically, the cage has a low electron affinity so that the
electrons stay well away from the wall.  This is reflected in the
localization of the wave function, and provides an intuitive
reason why a simple well is effectively realized.

If we further dope the system to have K$_3$LTA (which is experimentally 
in the magnetic 
regime), we have now four K atoms (the doped three on top of 
a red ball in Fig.1(b)), which form a cluster in the $\alpha$ cage. 
The precise atomic configuration has not been experimentally determined,
so we have focused here on the configuration where the four atoms 
(at $(0.5\pm x_1, 0.5\pm x_1, 0.5)$) form 
a square, with $x_1\simeq 0.25$ minimizing the total energy, although we 
have calculated for other possibilities such as a tetrahedral configuration. 

If we look at the band structure and the wave functions
for the K$_3$LTA in Fig.3, we can see that three bands around $E_F$ 
(located above the $s$ band that has fallen below $E_F$) have amplitudes within
the $\alpha$ cage. We identify these as $p_x$, $p_y$, and $p_z$
bands, respectively, as confirmed by a fit of the dispersions to
the tight-binding model, where the fit is again excellent.  
To be precise $E_F$ 
intersects the $p_x$ and $p_z$ bands, which are 
degenerate, reflecting the symmetry of the cluster.
The fitted hopping integrals for $s$, $p_x$, $p_y$ and $p_z$ bands are, 
respectively, $(t_x, t_y, t_z) \simeq (-30.0, -25.0, -0.5), 
(125, -25.0, -62.5), (-12.5, 25.0, -50.0)$ and $(-12.5, -0.5, 75.0)$ 
in meV. 
The hopping integrals are almost an order of magnitude greater than 
those for K$_1$LTA, as expected from the larger cluster size.  
An ESR experiment shows that the $g$ value decreases for K$_n$LTA with 
$n>2$\cite{Nakano01}. This may be understood as the 
degeneracy of the $p$ bands enhancing the spin-orbit
interaction and reducing the $g$ value.

Now we come to a big question of whether 
the system is strongly correlated.  As mentioned, 
a puzzle is that experimentally K$_1$LTA and K$_3$LTA are 
insulators, while the LDA finds them to be metals.  
So we first estimate the Coulomb matrix elements. The largest one 
is the intra-orbital Coulomb interaction $U$, which is 
$U=\int |\phi_1({\bf x})|^2 V({\bf x}-{\bf y})|
\phi_1({\bf y})|^2 d{\bf x}d{\bf y}$, 
where $\phi$'s are the wave functions at $\Gamma$.  
Here  $V$ is assumed to be the bare 
Coulomb interaction as a first approximation, since 
the $s$ and $p$ wave functions are well localized in the interior
of each cage\cite{AddComm}.
$U$ is calculated to be $U\simeq 4.5$ eV for the $s$ band in
K$_1$LTA, and $\simeq 4.0$ eV for the $p_x$ band in K$_3$LTA.
Given that $U/W \sim 10\gg 1$, where $W$ is the band
width, we can expect that these materials are Mott insulators.

However, since the relevant bands are $p$ bands with very anisotropic 
dispersions, we have to be careful in estimating 
the critical $U_c$ for the metal-insulator 
transition.  Here we have 
employed the dynamical mean-field theory\cite{Georges96} with the maximum 
entropy method\cite{Jarrell96} to estimate the transition point 
by calculating the spectral function in the single-band Hubbard 
model\cite{kawakami} for a typically anisotropic ($t_x : t_y : t_z=$5:1:1) 
case as well as the isotropic (1:1:1) one.

Fig. 4 shows the spectra for various values of Hubbard $U/W$,
where we can see that the system becomes an insulator (as
identified from a gap in the spectral function) for $U/W$ larger
than $\simeq 2$\cite{commentB} in the anisotropic case.  So we
conclude that this particular K-doped zeolite is on the
Mott-insulator side.  In more general terms 
we expect that the metal-insulator
transition can be controllable through
control of $U/W$.  For example, a zeolite called faujasite is
known to be metallic when alkali-metal doped\cite{fau}, where
this form of zeolite has a significantly wider ($7$\AA\ against $5$\AA\ for
LTA) window between the cages, which should result in a smaller $U/W$.

We finally come to the magnetic property.  To discuss this we require 
the inter-orbital $U'$ and exchange integrals $J$
as well, where 
$U'=\int |\phi_1({\bf x})|^2 V({\bf x}-{\bf y})|
\phi_2({\bf y})|^2 d{\bf x}d{\bf y}$,
$J=\int \phi_1^\dagger({\bf x})\phi_2^\dagger({\bf y})
V({\bf x}-{\bf y})\phi_1({\bf y})\phi_2({\bf x}) 
d{\bf x}d{\bf y}$.  
For the $p_x$-$p_z$ pair in K$_3$LTA, we find
$U' \simeq 3.7$~eV, and $J \simeq 0.7$~eV, which are 
similar to those roughly estimated by Nozue {\it et al}\cite{Nozue93}.
Multi-orbital systems are in general favorable for ferromagnetism, 
since the inter-orbital kinetic-exchange coupling,
$J^{\rm inter}_{\rm F} = -2 t^2/(U'-J)$, is ferromagnetic 
(accompanied by an orbital superlattice structure).  $J^{\rm inter}_{\rm F}$ 
is estimated here to be 
$(40 \sim 100)$ K for $p_x$ and $p_z$ bands in K$_3$LTA.  
This coupling competes with the intra-orbital kinetic-exchange coupling, 
$J^{\rm intra}_{\rm AF} \simeq 4 Ut^2/(U^2-J^2)$, which is 
antiferromagnetic.  These exchange energies are 
$(60\sim 160)$ K for K$_3$LTA, which are two orders of 
magnitude greater than the exchange energy, $4t^2/U\simeq O(1)$ K, 
for the (single-band case of) K$_1$LTA, where the difference is 
mainly due to an order of magnitude difference in $t$.  
If we look at the magnetic phase diagram in the 
literature\cite{kusakabe,momoi},
K$_3$LTA is, as indicated in the inset of Fig.4,
right around the ferromagnetic phase boundary.

Experimentally, ferromagnetism with finite magnetization 
has been observed for K$_n$LTA with $2<n<7$\cite{Nozue92,Nozue93}, 
where the Curie temperature $\sim 10$ K
for $n\simeq 3$ while the spin susceptibility for $T>30$ K 
exhibits a Curie-Weiss law, $1/(T-\Theta)$, 
with the Weiss temperature $\Theta \simeq 0$ K
for K$_1$LTA, while $(-40 \sim -30)$ K for K$_3$LTA\cite{Nozue92,Nozue93}.
The doping dependence as well as the energy scale are 
consistent with the exchange interactions estimated here.  

Finally we comment on the Mott transition, for which 
K$_n$LTA is experimentally\cite{partialmag} insulating even when the 
nominal doping level (averaged $n$) is fractional.  
This may possibly be related to a coexistence of differently 
doped regions with a domain structure\cite{partialmag,commentbandwidth}.
It would be interesting to know whether 
superconductivity as in the high-$T_c$ cuprates can appear 
when we realize doped Mott insulators in the present system (by, e.g., 
degrading such a domain structure).  
Future work should also include 
an elaboration of the LDF+DMFT approach.  
These will enable us to systematically study electronic and magnetic 
effects in the ``supercrystal", for which some 
experimental and theoretical attempts are under way.

This work was supported by a Grant-in-Aid for Scientific Research (A) Fund 
from the Ministry of Education, Culture, Sports, Science 
and Technology, Japan.

\begin{figure}
\begin{center}
\leavevmode\epsfysize=40mm \epsfbox{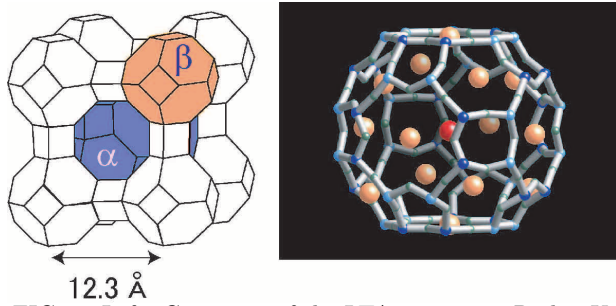}
\caption{
Left: Geometry of the LTA structure. Right: Unit cell 
(i.e., $\alpha$ cage) of the undoped zeolite considered here, 
with dark blue: Si, light blue: Al, dark green: oxygen, orange and red:
 K. 
}
\end{center}
\end{figure}

\begin{figure}
\begin{center}
\leavevmode\epsfysize=60mm \epsfbox{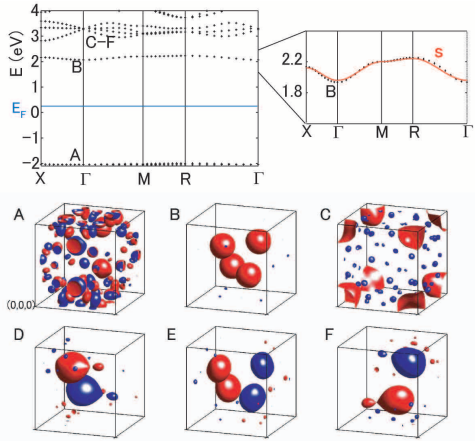}
\caption{
The band structure of the undoped zeolite LTA. The top right panel is 
a blowup of the lowest conduction band, where the red line is a tight-binding 
fit. Bottom panels are wave functions (red contours: positive, blue: negative) 
at $\Gamma$ in the bands A-F as labeled in the band structure.
}
\end{center}
\end{figure}

\begin{figure}
\begin{center}
\leavevmode\epsfysize=60mm \epsfbox{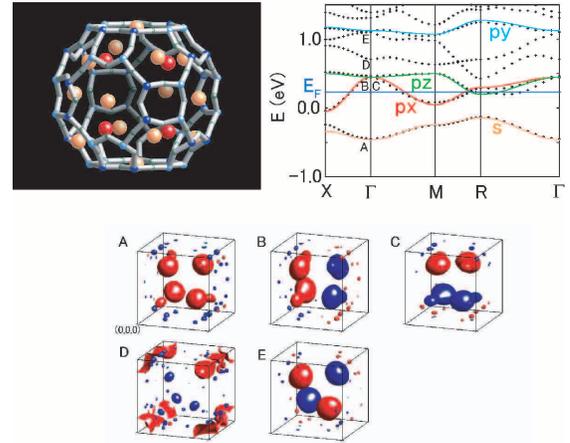}
\caption{
Band structure of a K-doped zeolite K$_3$LTA. Colored curves are a 
tight-binding fit. Top left inset depicts the atomic configuration, 
where four red K's (the doped K + the red K in Fig.1)
form a square in the $\alpha$ cage. Bottom panels are wave functions at 
$\Gamma$ in the bands A-E as labeled in the band structure.
}
\end{center}
\end{figure}

\begin{figure}
\begin{center}
\leavevmode\epsfysize=50mm \epsfbox{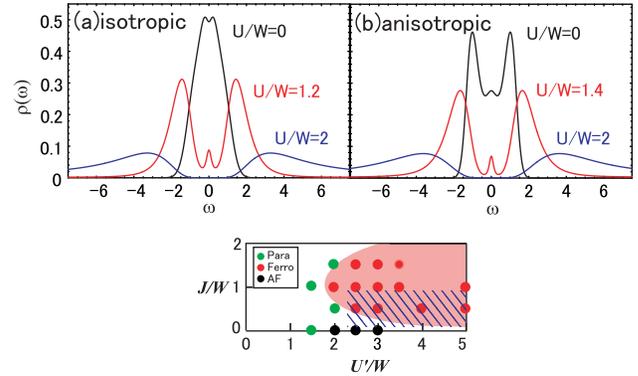}
\caption{
Spectral function, $\rho(\omega)$, obtained with the dynamical mean-field 
theory for the Hubbard model on the cubic lattice with the isotropic(a) or 
an anisotropic(b) dispersions for various values of $U/W$. 
The bottom inset shows the magnetic phase diagram for 
the degenerate Hubbard model against the inter-orbital Coulomb ($U'$) 
and the Hund coupling ($J$) interactions 
after Ref.\protect\cite{momoi}, where the parameter region for the doped 
zeolite is indicated by blue hatch with its widths representing 
the dependence on the atomic configuration of the cluster 
and on the assumption of $U=U'+2J$ on which the diagram is drawn. 
$W$ is here defined as the width of the gaussian density of states.
}
\end{center}
\end{figure}

\end{multicols}

\begin{references}
\bibitem{AddComm}The interaction parameters calculated 
by the Slater integrals can be
overestimated, by $> 1.3$ for isolated ions and $> 2$ for solids, but
the cluster in each cage here is closer to the former, and even if the
factor is $\sim 2$, this will not alter the present conclusion
on the Mott-insulator and magnetic properties.
\bibitem{commentB} For an infinite-dimensional system the metal-insulator 
transition point has been estimated to be 
$U/W = 1 \sim 2$\cite{Georges96}.
\bibitem{commentbandwidth} 
A super-structure in the array of clusters 
is experimentally detected for $n > 2$.
\end{references}
\end{document}